\documentclass[12pt]{article}
\usepackage{graphicx}

\begin{document}

\title{Strangeness and hypernuclear production in fragmentation reactions induced by antikaons}

\author{Zhao-Qing Feng \footnote{Corresponding author. \newline \emph{E-mail address:} fengzhq@scut.edu.cn}}
\date{}
\maketitle

\begin{center}
\small \emph{School of Physics and Optoelectronics, South China University of Technology, Guangzhou 510640, China}
\end{center}

\textbf{Abstract}
\par
Formation mechanism of hyperfragments with strangeness s=-1 and s=-2 in collisions of antikaons on nuclei has been investigated within a microscopic transport model. Dynamics of pseudoscalar mesons and hyperons is modeled within the transport model, in which all possible reaction channels for creating hyperons such as the elastic scattering, resonance production and decay, strangeness exchange reaction and direct strangeness production in meson-baryon and baryon-baryon collisions have been included. A coalescence approach is developed for constructing hyperfragments in phase space and the decay process is described with a statistical approach. It is found that the $\Xi^{-}$ production is correlated to the K$^{+}$ formation and the hyperons $\Lambda$ and $\Sigma$ are created within a broad rapidity region. The production cross sections of nucleonic fragments and hyperfragments weakly depends on the incident momentum. The yields of $\Lambda-$ hyperfragments are the six order of magnitude of $\Xi^{-}-$ hyperfragments.

\bigskip

Inclusion of the strangeness degree of freedom in nuclear medium extends the research activities in nuclear physics, in particular on the issues of the nuclear structure of hypernucleus and kaonic nucleus, hyperon-nucleon and hyperon-hyperon interactions, probing the in-medium properties of hadrons \cite{Gi95, Ha06}. Moreover, hadrons with strangeness as essential ingredients influence the high-density nuclear equation of state (EOS). The strangeness constitution in dense matter softens the EoS at high-baryon densities, and consequently increases the mass of neutron stars\cite{Ji13,We12}. Since the first observation of $\Lambda$-hypernuclide in nuclear multifragmentation reactions induced by cosmic rays in 1950s \cite{Da53}, a remarkable progress has been obtained in producing hypernuclides via different reaction mechanism, such as hadron (pion, K$^{\pm}$, proton, antiproton) induced reactions, bombarding the atomic nucleus with high-energy photons or electrons, and fragmentation reactions with high energy heavy-ion collisions. Experimental collaborations of nuclear physics facilities in the world have started or planned to investigate hypernuclei and their properties, e.g., PANDA \cite{Pand}, FOPI/CBM and Super-FRS/NUSTAR \cite{Sa12} at FAIR (GSI, Germany), STAR at RHIC (BNL,USA) \cite{Star}, ALICE at LHC (CERN) \cite{Do13}, NICA (Dubna, Russia) \cite{Nica}, J-PARC (Japan) \cite{Ta12}, HIAF (IMP, China) \cite{Ya13}. In these laboratories, the strangeness nuclear physics is to be concentrated on the isospin degree of freedom (neutron-rich/proton-rich hypernuclei), multiple strangeness nucleus, anti hypernucleus, high-density hadronic matter with strangeness. The hypernuclear spectroscopy and kinematics in the antikaon induced reactions have been investigated in experiments \cite{Na14,Ga16}. Theoretical description for hypernucleus formation is helpful for managing the detector system in experiments.

The dynamics mechanism of hypernucleus formation in the antikaon induced reactions is a complex process and related to the creation of hyperons, propagation, hyperfragment construction and decay. Up to now, several models have been established for describing the hypernucleus formation in nuclear reactions, i.e., the Statistical Multifragmentation Model (SMM) \cite{Bo07}, statistical approach with a thermal source \cite{An11} and microscopic transport model \cite{Bo15}. Some interesting results are obtained for understanding the hypernucleus production, i.e., the yields of hyperfragments, fragment production with multiple strangeness, dibaryon states etc.

In this work, the strangeness production and hypernuclear dynamics in the antikaon induced reactions are to be investigated within the Lanzhou quantum molecular dynamics (LQMD) transport model. In the model, the dynamics of resonances ($\Delta$(1232), N*(1440), N*(1535) etc), hyperons ($\Lambda$, $\Sigma$, $\Xi$) and mesons ($\pi$, $K$, $\eta$, $\overline{K}$, $\rho$, $\omega$ ) is included and coupled to the reaction channels via hadron-hadron collisions and decay of resonances \cite{Fe11,Fe18}. The evolutions of hadrons in nuclear medium are described by Hamilton's equations of motion under the self-consistently generated mean-field potentials. The Hamiltonian of mesons and hyperons is constructed as follows
\begin{eqnarray}
H = \sum_{i=1}^{N_{H}}\left( V_{i}^{\textrm{Coul}} + \omega(\textbf{p}_{i},\rho_{i}) \right).
\end{eqnarray}
Here, $N_{P}$ is the total number of mesons or hyperons. The hyperon mean-field potential is constructed on the basis of the light-quark counting rule. The self-energies of $\Lambda$ and $\Sigma$ are assumed to be two thirds of that experienced by nucleons. And the $\Xi$ self-energy is one third of nucleon's ones. Thus, the in-medium dispersion relation reads
\begin{equation}
\omega(\textbf{p}_{i},\rho_{i})=\sqrt{(m_{Y}+\Sigma_{S}^{Y})^{2}+\textbf{p}_{i}^{2}} + \Sigma_{V}^{Y},
\end{equation}
e.g., for hyperons $\Sigma_{S}^{\Lambda,\Sigma}= 2 \Sigma_{S}^{N}/3$, $\Sigma_{V}^{\Lambda,\Sigma}= 2 \Sigma_{V}^{N}/3$, $\Sigma_{S}^{\Xi}= \Sigma_{S}^{N}/3$ and $\Sigma_{V}^{\Xi}= \Sigma_{V}^{N}/3$. The nucleon scalar $\Sigma_{S}^{N}$ and vector $\Sigma_{V}^{N}$ self-energies are computed from the well-known relativistic mean-field model with the NL3 parameter ($g_{\sigma N}$=8.99, $g_{\omega N}$=12.45 and $g_{\rho N}$=4.47). The optical potential of hyperon is derived from the in-medium energy as $V_{opt}(\textbf{p},\rho)=\omega_{Y}(\textbf{p},\rho)-\sqrt{\textbf{p}^{2}+m^{2}}$. The values of optical potentials at saturation density are -32 MeV and -16 MeV for $\Lambda(\Sigma)$ and $\Xi$, respectively. The attractive potential is available for the bound hyperfragment formation. The in-medium effects of pseudoscalar mesons ($\pi$, $\eta$, $K$ and $\overline{K}$ ) in heavy-ion collisions have been investigated and the optical potentials at the saturation density were extracted \cite{Fe18,Fe15}.

In the K$^{-}$ induced reactions, hyperons are mainly created via the strangeness exchange reaction of $K^{-}N \rightarrow  \pi Y$ and the direct channel of $K^{-}N \rightarrow K^{+}\Xi$. The cross section of $K^{-}N \rightarrow  \pi Y$ is implemented by fitting the available experimental data \cite{Fe18}. The $\Xi$ production is estimated by the parametrized formula \cite{Li12}, which is basically consistent with the calculations by a phenomenological model \cite{Sh11}. The isotropic distribution of hyperon is assumed once it created. The yields of $\Lambda$ and $\Sigma$ are abundant because of larger cross sections. The production rate of $\Xi$ is low, but can be easily captured by  nucleonic fragments to form hyperfragments owing to the low relative momentum.

The target nuclide can be heated by incoming energetic antikaon, which leads to the fragmentation reaction. The primary fragments are constructed in phase space with a coalescence model, in which nucleons at freeze-out are considered to belong to one cluster with the relative momentum smaller than $P_{0}$ and with the relative distance smaller than $R_{0}$ (here $P_{0}$ = 200 MeV/c and $R_{0}$ = 3 fm). The hyperons are captured by nucleonic fragments to form $\Lambda$-fragments. Here, a larger relative distance ($R_{0}$ = 5 fm) and the relative momentum similar to nucleonic ones ($P_{0}$ = 200 MeV/c) between hyperon and nucleon in constructing a hypernucleus, which is caused from the fact that the weakly bound of hypernucleus with a bigger rms (root-mean-square) radius, e.g., 5 fm rms for $^{3}_{\Lambda}$H and 1.74 fm for $^{3}$He \cite{Ar04}. At the freeze-out, the primary hyperfragments are highly excited. The de-excitation of the hyperfragments is described within the GEMINI code \cite{Ch88}, in which the decay width of light fragments with Z$\leq$2 and the binary decay are calculated by the Hauser-Feshbach formalism \cite{Ha52} and transition state formalism \cite{Mo75}, respectively. We implemented the hyperon decay based on the Hauser-Feshbach approach. The binding energy of hyperon is evaluated by a phenomenological formula \cite{Sa06}. Shown in Fig. 1 is the decay widths of neutron, proton, $\alpha$, $\Lambda$ and fission from excited $^{40}_{\Lambda}$Ca and $^{238}_{\Lambda}$U. The $\Lambda$ emission is the dominant decay mode for the light nucleus because of small separation energy. The neutron evaporation and binary fission are competitive for heavy nucleus.

The emission mechanism of particles produced in antikaon induced reactions is significant in understanding contributions of different reaction channels associated with antikaons on nucleons and secondary collisions. Shown in Fig. 2 is the excitation functions of $\pi$, $\eta$, K$^{+}$, $\Lambda$, $\Sigma$ and $\Xi^{-}$ in the reaction of K$^{-}$+$^{40}$Ca. It is obvious that the $\pi$ and hyperons ($\Lambda$, $\Sigma$) slightly decrease with the beam momentum. The K$^{+}$ and $\Xi^{-}$ can be created above the threshold energy ($p_{th}=$1.04 GeV/c) and the yields increase rapidly with the momentum. Once the low-momentum hyperon is created, it can be easily captured by the potential well to form hyperfragments. Both the abundance and phase-space distribution of produced hyperons contribute the production rate of hypernuclide.

The hyperon production in heavy-ion collisions near threshold energies has been investigated extensively in experiments and in theories, in particular on the issues of nuclear equation of state, hyperon-nucleon interaction, correlation of hyperon production etc \cite{St01,Zh04,Ha12,Fe13}. The hyperon distribution in phase space dominates the hypernuclear formation in the antikaon induced reactions. Shown in Fig. 3 is the rapidity distribution of hyperons produced in the reactions of K$^{-}$ on $^{12}$C, $^{20}$Ne, $^{40}$Ca, $^{95}$Mo, $^{124}$Sn, $^{165}$Ho, $^{197}$Au and $^{238}$U at the incident momentum of 1.5 GeV/c. A broad structure is obvious for the production of $\Lambda$ and $\Sigma$ with enough cross sections. The $\Xi^{-}$ yields are strongly reduced with the narrow shape. The invariant spectra of particle production manifest the hadronic matter properties. We calculated the kinetic energy spectra of $\pi^{-}$, K$^{+}$, $\Lambda+\Sigma^{0}$ and $\Xi^{-}$ as shown in Fig. 4. A steep structure is pronounced for the K$^{+}$ production. The reabsorption process by surrounding nucleons retards the particle emission and leads to a platform in the energy spectra.

The nuclear dynamics induced by antikaons is motivated on the aspects of the strangeness exchange reactions, nuclear fragmentation, liquid-gas phase transition, hypernuclide formation etc. On the other hand, the antikaon-nucleus collisions have advantages to investigate the energy dissipation mechanism, the interaction of strange particle and nucleon, hadronic matter properties around the saturation density. Shown in Fig. 5 is a comparison of the nucleonic fragments, $\Lambda$-hyperfragments and $\Xi^{-}$- hyperfragments in the $K^{-}$+$^{40}$Ca reaction at different incident momenta. The strangeness exchange reaction of $K^{-}N \rightarrow \pi  Y$ dominates the hyperon production. Once the $\Lambda$ is created in the nucleus, a large probability is captured to form a hypernucleus. The nucleonic fragments are mainly produced in the target-like region and the light clusters are from the decay of excited fragments. The fragmentation in the antikaon induced reactions weakly depends on the incident energy. The production of $\Lambda$-hyperfragments is roughly the six-order magnitude of the $\Xi^{-}$-hypernuclear yields. The antikaon induced reactions are favorable for producing the target-like hyperfragments, in particular for creating heavy hypernuclides in comparison to heavy-ion collisions. The production cross sections are measurable in experiments, i.e., the $\Lambda$-hypernuclide with the magnitude of mb and the double strangeness hypernuclear formation at the level of nb. The mass and charge spectra of nucleonic fragments and hyperfragments in the fragmentation reaction on $^{95}$Mo is calculated as shown in Fig. 6. The hypernuclear production is reduced the 3-order magnitude of nucleonic fragments.

In summary, the particle production in K$^{-}$ induced reactions has been investigated within the LQMD transport model. The $\pi$, $\Lambda$ and $\Sigma$ are mainly created from the strangeness exchange reactions. The abundance of K$^{+}$ and $\Xi^{-}$ increases rapidly with the incident momentum. The broad rapidity and transverse momentum spectra are obtained for $\Lambda$ and $\Sigma$ production, which enable the formation of hyperfragments. The fragmentation takes place in the target-like region. The double strangeness hypernuclides in K$^{-}$ induced reactions are formed with the cross section at the level of nb, which are independent on the incident momentum once above the threshold energy of $\Xi^{-}$ production.

\textbf{Acknowledgements}
This work was supported by the National Natural Science Foundation of China (Projects No. 11722546 and No. 11675226) and the Talent Program of South China University of Technology.

\newpage

\begin{figure}
\begin{center}
{\includegraphics*[width=0.8\textwidth]{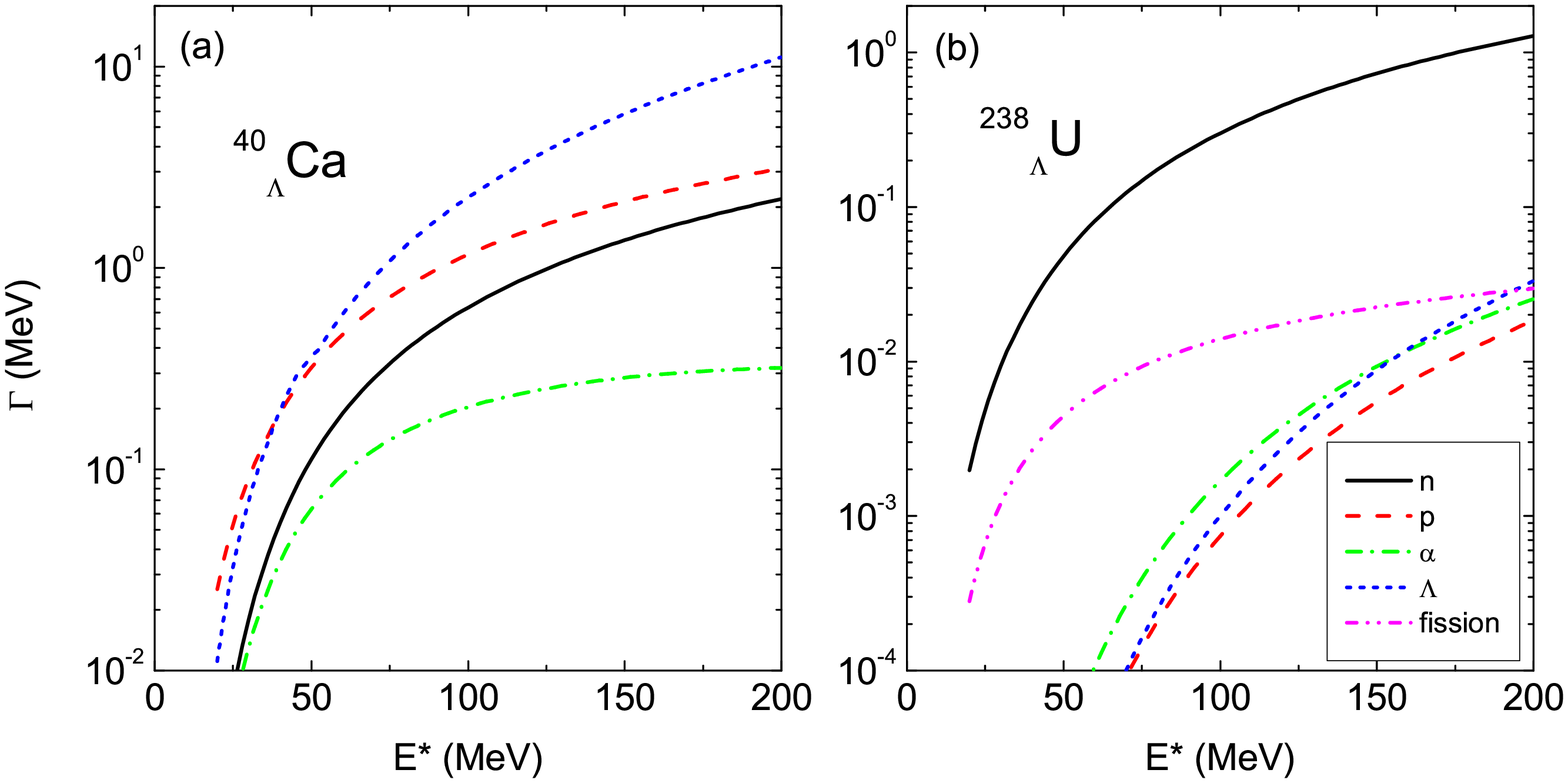}}
\end{center}
\caption{Decay width for evaporating neutron, proton, $\alpha$, $\Lambda$ and fission of $^{40}_{\Lambda}$Ca and $^{238}_{\Lambda}$U as a function excitation energy.}
\end{figure}

\begin{figure}
\begin{center}
{\includegraphics*[width=0.8\textwidth]{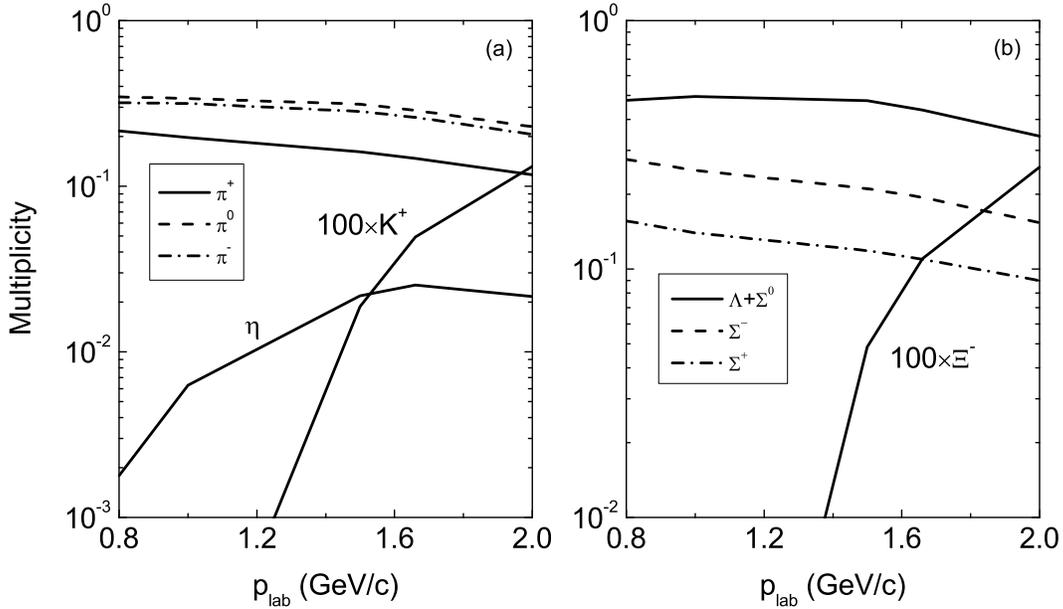}}
\end{center}
\caption{Production of mesons and hyperons in K$^{-}$ induced reactions on $^{40}$Ca as a function of beam momentum.}
\end{figure}

\begin{figure}
\begin{center}
{\includegraphics*[width=0.8\textwidth]{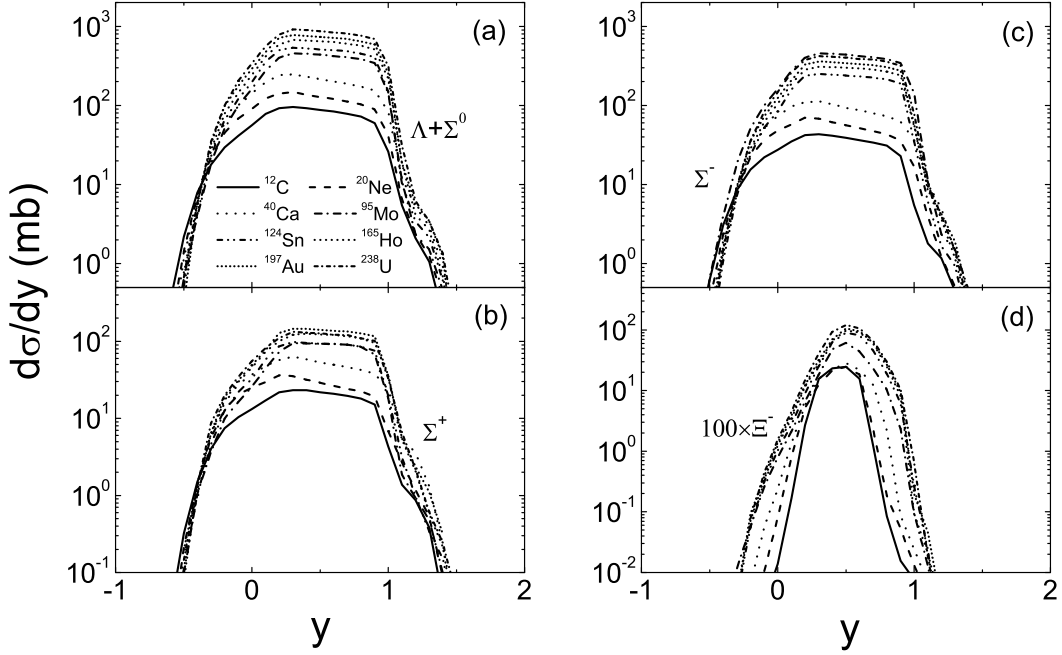}}
\end{center}
\caption{Rapidity distribution of free hyperons in collisions of K$^{-}$ on $^{12}$C, $^{20}$Ne, $^{40}$Ca, $^{95}$Mo, $^{124}$Sn, $^{165}$Ho, $^{197}$Au and $^{238}$U at the incident momentum of 1.5 GeV/c.}
\end{figure}

\begin{figure}
\begin{center}
{\includegraphics*[width=0.8\textwidth]{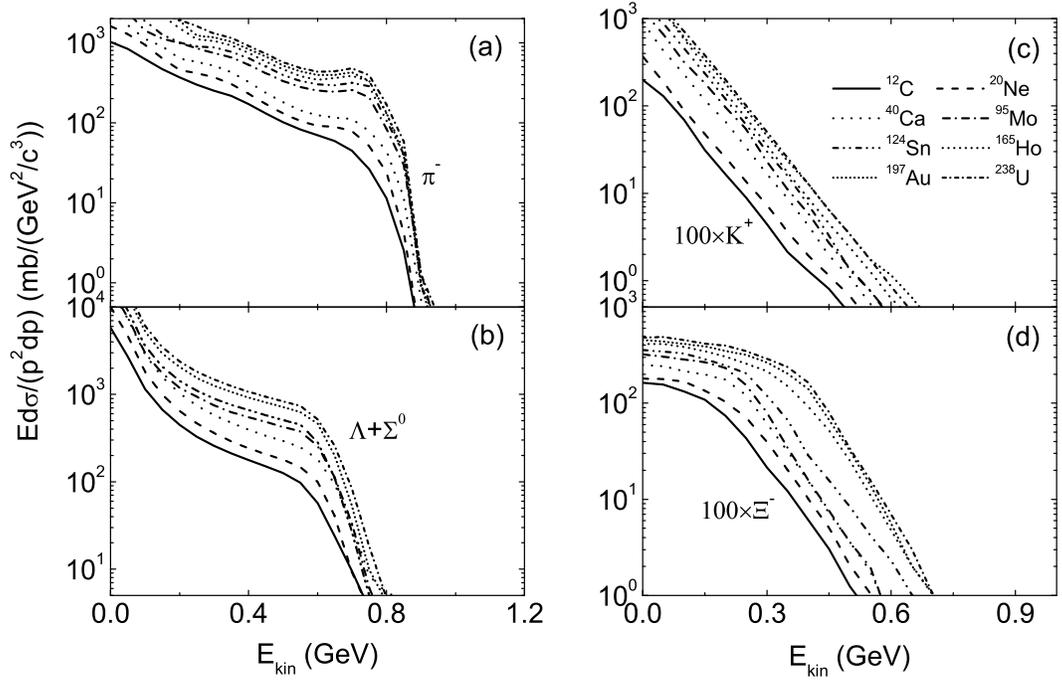}}
\end{center}
\caption{Inclusive spectra of $\pi^{-}$, K$^{+}$, $\Lambda+\Sigma^{0}$ and $\Xi^{-}$ in the K$^{-}$ induced reactions on different target nuclei at the momentum of 1.5 GeV/c.}
\end{figure}

\begin{figure}
\begin{center}
{\includegraphics*[width=0.8\textwidth]{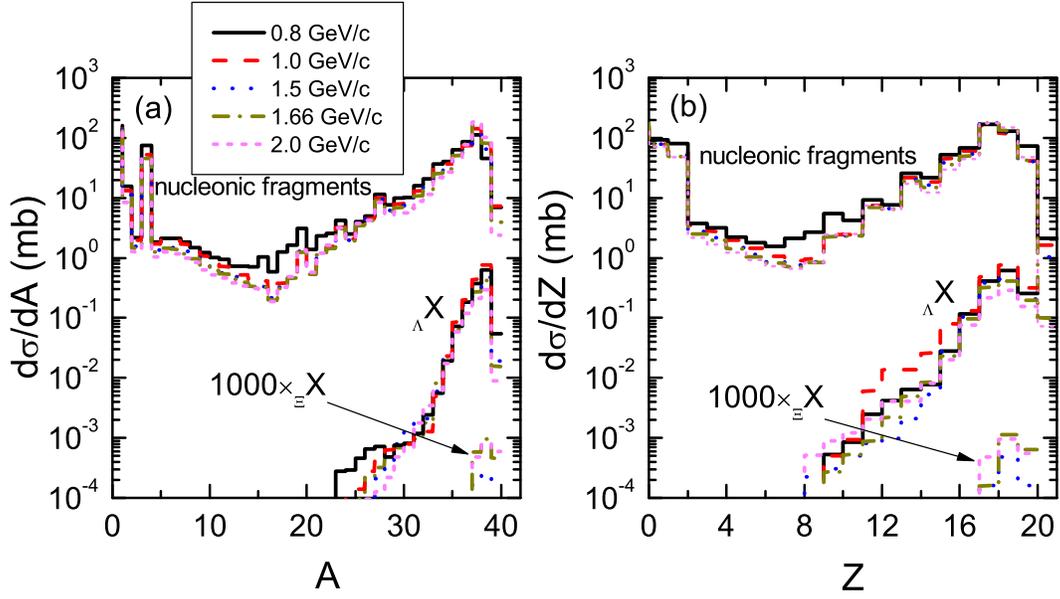}}
\end{center}
\caption{Incident energy dependence of the nucleonic fragments, $\Lambda-$ hyperfragments and $\Xi^{-}-$ hyperfragments as functions of mass and charged numbers in collisions of K$^{-}$+$^{40}$Ca, respectively.}
\end{figure}

\begin{figure}
\begin{center}
{\includegraphics*[width=0.8\textwidth]{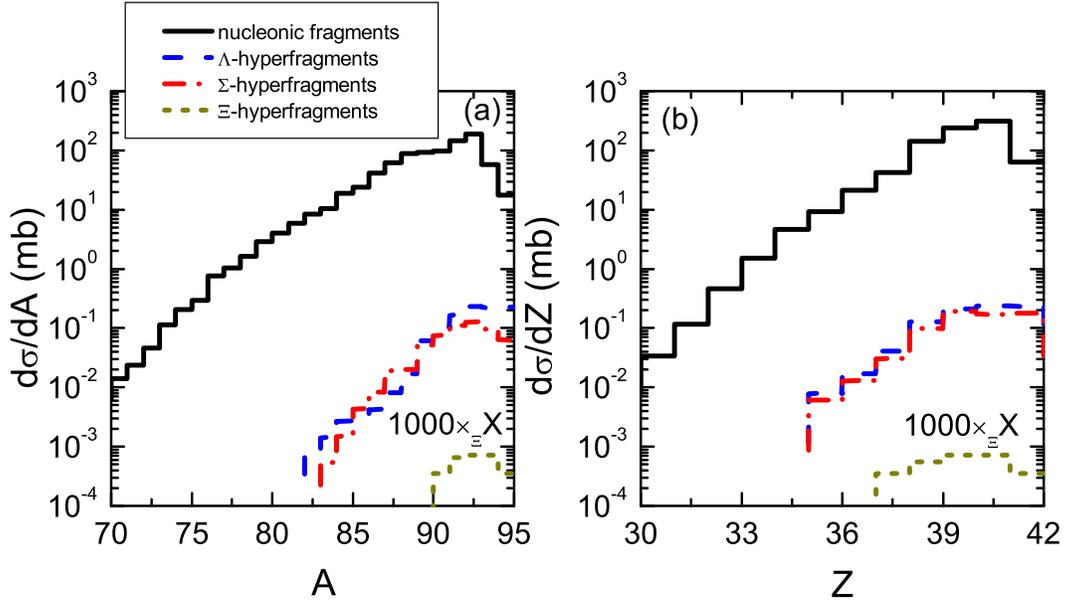}}
\end{center}
\caption{Comparison of the nucleonic fragments, $\Lambda-$, $\Sigma-$ and $\Xi^{-}-$hyperfragments in the K$^{-}$ induced reactions on $^{95}$Mo at the incident momentum of 1.5 GeV/c.}
\end{figure}

\end{document}